# Tissue-specific predictive performance: A unified estimation and inference framework for multi-category screening tests

A. Gregory DiRienzo*, Elie Massaad & Hutan Ashrafian

Harbinger Health, Cambridge, MA, USA

* Corresponding author email: gdirienzo@harbinger-health.com

**Abstract**

Multi-Cancer Early Detection (MCED) testing with tissue localization aims to detect and identify multiple cancer types from a single blood sample. Such tests have the potential to aid clinical decisions and significantly improve health outcomes. Despite this promise, MCED testing has not yet achieved regulatory approval, reimbursement or broad clinical adoption. One major reason for this shortcoming is uncertainty about test performance resulting from the reporting of clinically obtuse metrics. Traditionally, MCED tests report aggregate measures of test performance, disregarding cancer type, that obscure biological variability and underlying differences in the test's behavior, limiting insight into true effectiveness. Clinically informative evaluation of an MCED test's performance requires metrics that are specific to cancer types. In the context of a case-control sampling design, this paper derives analytical methods that estimate cancer-specific intrinsic accuracy, tissue localization readout-specific predictive value and the marginal test classification distribution, each with corresponding confidence interval formulae. A simulation study is presented that evaluates performance of the proposed methodology and provides guidance for implementation. An application to a published MCED test dataset is given. These statistical approaches allow for estimation and inference for the pointed metric of an MCED test that allow its evaluation to support a potential role in early cancer detection. This framework enables more precise clinical decision-making, supports optimized trial designs across classical, digital, AI-driven, and hybrid stratified diagnostic screening platforms, and facilitates informed healthcare decisions by clinicians, policymakers, regulators, scientists, and patients.

## 1. Introduction

Multi-cancer early detection (MCED) tests use a single blood sample to ascertain the presence of multiple types of cancers (1). These tests inherently accommodate diverse populations of individuals composed of multiple clinical disease categories and attempt to classify individuals into one of multiple cancer types or a healthy state, resulting in a multi-dimensional confusion matrix rather than a conventional two-by-two table. This larger dimensionality results in a more complex evaluation of test performance compared to single-cancer binary tests. Conventional binary tests typically consider a two-way classification of disease versus no disease that leads to straightforward analyses of test performance (sensitivity, specificity, positive and negative predictive value). Deriving an appropriate analysis framework for MCED tests presents challenges for both defining clinically meaningful and actionable metrics and deriving valid estimation and inference procedures for such metrics.

Performance metrics for used to evaluate conventional binary tests that focus on the presence or absence of one disease allow a straight-forward clinically-focused interpretation. On the contrary, the broader applicability of MCED tests and their simultaneous assessment and categorization of multiple disease states does not afford a clear and obvious formulation of clinically meaningful and actionable performance metrics. Furthermore, the development of statistical estimation and inference techniques for such metrics requires careful consideration of the underlying nuances in the data that are revealed upon this dimensionality expansion. Specifically, the increased dimensionality corresponding to MCED tests complicates:

i. the identification and definition of clinically meaningful performance metrics that requires accommodation of the potential for varying performance across cancer types, differences in cancer incidence rates, and dynamic clinical impacts;
ii. statistical estimation and inference procedures of such metrics, due to increased complexity in appropriately quantifying uncertainty, the need for accommodating sparse data and accounting for covariance among empirical cancer-specific performance metrics;
iii. clinical interpretation and decision-making, since translating these complex multidimensional analytical frameworks into straightforward and actionable clinical insights for practitioners, policymakers, and regulatory agencies becomes significantly more challenging.

The purpose of this paper is to propose a unified framework that identifies and defines clinically relevant performance metrics for MCED tests along with analytical formulas for unbiased estimation and valid confidence interval construction in the context of a case-control study design. Evidence generation from case-control studies conducted in the intended use population is a necessary component of evaluating the performance of screening and diagnostic tests. At least two important reasons are:

i. compared to prospective cohort studies that enroll participants with unknown clinical outcome status, the procurement of case samples is structurally independent from the number of healthy individuals that otherwise need to be enrolled to achieve a target number of cases based on an expected disease incidence and potentially lengthy clinical follow-up period.

ii. unlike prospective cohort studies typically only perform follow-up workup to establish the clinical outcome for those subjects that are a test positive at enrollment, case-control studies allow direct estimation of and inference for specificity and intrinsic accuracy metrics.

However, because cases are oversampled by design, case-control designs do not provide information on disease incidence, thereby precluding direct estimation of predictive value-type metrics and the marginal test readout distribution. We derive estimation and inference procedures for such metrics in this setting using two approaches that provide information on incidence.

After completion of one or more case-control studies, viable screening and diagnostic tools will generally have performance quantified using long-term and expensive prospective cohort studies conducted in the intended use population. The information obtained from case-control studies can have great value not only for aiding the design of future prospective studies but also impacting clinical practice. For example, the result of the MCED test is generally intended to provide the practicing physician additional information to use as a part of the clinical decision-making process for first-line procedures. In particular, quantifying the MCED test's ability to correctly indicate the appropriate follow-up action on an individual and the corresponding contribution to ultimately improve outcomes in the intended use population is critical in assessing the test's clinical utility. This information is extremely valuable for patients, providers, regulators and payers before large and long-term prospective cohort studies can be completed.

Authors have advocated for the performance of MCED tests to be quantified specifically for each cancer type under consideration in datasets representative of the intended use population (2). Specifically, metrics that establish the intrinsic accuracy of each cancer type evaluated by the MCED test and the predictive value positive for each Tissue of Origin (TOO) test readout category are critical for assessing the test's clinical utility. Obtaining intrinsic accuracy metrics for each cancer type quantifies an MCED test's ability to control both false negative and incorrect TOO classification results, while predictive value metrics for each TOO readout category address the corresponding false positive error and associated risk of incorrectly informing decisions of ensuing clinical follow-up.

The proposed analysis framework accommodates two approaches to inform disease incidence. When the intended use population is represented by that of a national registry or published large population study, incidence rates from such databases can be employed. Alternatively, in settings where incidence information is not available from external sources, a sample-derived approach for obtaining conditional disease-type incidence that is paired with a posited overall value for disease incidence is proposed; this value can be varied in a sensitivity analysis if warranted.

Section 2 briefly reviews standard metrics for 2 x 2 tables and Section 3 expands these definitions to derive clinically meaningful metrics corresponding to a *(J+1)* x *(K+1)* joint distribution of disease state versus MCED test readout. In Section 4, unbiased estimation and valid inference for various types of desired metrics is provided in the context of a case-control sampling design. First, estimators for intrinsic accuracy for a given cancer state are derived that recognize these quantities are in fact compound random variables because, although the total number of cases is predetermined, generally, the realized number of cases of a specific cancer state varies randomly. In addition, metrics are defined for the false-negative probability for each cancer state as well as an overall intrinsic accuracy metric. Estimation and inference procedures are also provided for predictive value negative and, separately for each TOO readout category, predictive value positive; a method for obtaining overall predictive value positive is also proposed. Estimation and inference for the marginal test readout category distribution is presented; this can be useful for evaluating the impact of the MCED test on outcomes in the intended use population and for aiding the design of future prospective studies, for example informing the number of subjects needed to be enrolled to realize an expected number of positive test readouts. An approach is proposed to adjust false-positive counts for tests with very high specificity to stabilize estimation and inference for predictive value metrics and the marginal test distribution in the presence of this sparse data. Inference for each of type of metric is performed by using the delta-method on the logit transform to obtain valid confidence intervals for the corresponding population parameter in the unit interval. Stratified analyses are discussed in Section 5 and Section 6 presents a numerical simulation study that evaluates performance of the proposed methodology in various settings. Analysis of a real-world case-control study is provided in Section 7 and Section 8 concludes with a discussion.

## 2. Standard metrics for binary tests and disease states using case-control sampling

With a binary test readout “+” or “-” and binary disease outcome (Disease (“Case”) or no Disease (“Control”)), the two-way joint distribution of test and outcome is:

| | Test - | Test + | TOTAL |
|---|---|---|---|
| **Control** | $n_{00}$ | $n_{01}$ | $N_0$ |
| **Case** | $n_{10}$ | $n_{11}$ | $N_1$ |
| **TOTAL** | $n_{+0}$ | $n_{+1}$ | $N$ |

**Table 1:** Standard two-way frequency distribution of test versus disease state.

A case-control design fixes in advance each the number of cases and controls that are included in the study; under a random sampling model for each group, a binomial sampling model applies for the number of cases that test positive and the number of controls that test positive. That is, $n_{01} \sim Binomial\{N_0, P(Test + | \; no \; Disease)\}$ and $n_{11} \sim Binomial\{N_1, P(Test + | \; Disease)\}$, where $P(X)$ is the probability measure of the random variable $X$ arising from the target intended use population, denoted by $\boldsymbol{\mathcal{P}}$.

The test performance metrics sensitivity and specificity are defined as:

$$\textit{sensitivity (SE)} = P(Test + \mid Disease)$$

$$\textit{specificity (SP)} = P(Test - \mid no \; Disease) = 1 - \; P(Test + \mid no \; Disease)$$

and are directly estimable from this sampling model. The corresponding maximum likelihood estimators are the binomial proportions $\widehat{SE} = n_{11}/n_{1+}$ and $\widehat{SP} = n_{00}/n_{0+}$, respectively.

The positive (PPV) and negative (NPV) predictive value of the test are not directly estimable with the case-control sampling model since cases are oversampled from the intended use population $\boldsymbol{\mathcal{P}}$. To obtain estimates of these metrics, *Bayes rule* and the *Law of total probability* need to be used along with a working value for the disease incidence, denoted $P(D)$. The disease incidence is the expected number of cases per 1 person-year of follow-up in individuals from $\boldsymbol{\mathcal{P}}$, with $1 - P(D) = P(\overline{D})$ since $D \cup \overline{D} = \boldsymbol{\mathcal{P}}$ and the events of Disease ($D$) and no Disease ($\overline{D}$) are mutually exclusive, that is, $D \cap \overline{D} = \{\emptyset\}$. Using Bayes' rule, PPV = $P(D \mid Test{+}) = P(Test + \mid D)P(D)/P(Test +) = SE(P(D))/P(Test{+})$. Using the Law of total probability, $P(Test +) = SE\big(P(D)\big) + (1 - SP)(1 - P(D))$. Similarly, NPV = $P(\overline{D} \mid Test{-}) = SP(1 - P(D))/\{SP(1 - P(D)) + (1 - SE)P(D)\}$. Methods exist for confidence interval construction for PPV and NPV in this 2 x 2 setting shown in Table 1 (3).

## 3. Metrics for multi-category tests and disease states

Suppose that there are $J > 2$ mutually exclusive disease states, labeled $D_j, j = 1, \ldots, J$; let $D_0$ denote the no disease state. Then $\cup_{j=0}^{J} D_j = \boldsymbol{\mathcal{P}}$, with $D_j \cap D_\ell = \{\emptyset\}, (j \neq \ell)$, and $\sum_{j=0}^{J} P\big(D_j\big) = 1$. Each disease state may include one or more distinct disease types that share a common implication regarding the test's clinical utility; for example, a group of disease types that share a common first-line clinical workup. In addition to a Negative category, the test has a readout category corresponding to *K* different disease states (TOOs), where each of the test readout categories are denoted $Test = k : k = 0, \ldots, K$, with $Test = 0$ denoting the Negative readout and test positive categories 1 to *K* assumed to be a one-to-one correspondence to the disease states $D_1$ to $D_K$; there may be disease states, labeled $D_{K+1}$ to $D_J$, that do not have a corresponding test readout category. It is assumed that the data arise as a random sample of $N_0$ subjects with no disease and $(N - N_0)$ cases from intended use population $\mathcal{P}$. Table 2 below shows a depiction of the joint distribution in this setting.

| | ***Test=0*** | ***Test=1*** | $\cdots$ | ***Test=K*** | **TOTAL** |
|---|---|---|---|---|---|
| $\boldsymbol{D_0}$ | $n_{00}$ | $n_{01}$ | $\cdots$ | $n_{0K}$ | $N_0$ |
| $\boldsymbol{D_1}$ | $n_{10}$ | $n_{11}$ | $\cdots$ | $n_{1K}$ | $n_{1+}$ |
| $\vdots$ | $\vdots$ | $\vdots$ | $\cdots$ | $\vdots$ | $\vdots$ |
| $\boldsymbol{D_K}$ | $n_{K0}$ | $n_{K1}$ | $\cdots$ | $n_{KK}$ | $n_{K+}$ |
| $\vdots$ | $\vdots$ | $\vdots$ | $\cdots$ | $\vdots$ | $\vdots$ |
| $\boldsymbol{D_J}$ | $n_{J0}$ | $n_{J1}$ | $\cdots$ | $n_{JK}$ | $n_{J+}$ |
| **TOTAL** | $n_{+0}$ | $n_{+1}$ | $\cdots$ | $n_{+K}$ | $N$ |

**Table 2:** General two-way frequency distribution of a multi-category test and disease states.

There are two general types of metrics that convey information regarding the relationship between test and disease. *Intrinsic accuracy* measures the conditional probability of the target test result among individuals with a given disease state; *Predictive value* measures the conditional probability of the target disease state among individuals with a given test category readout.

Intrinsic accuracy corresponding to disease state $j = 0, \ldots, K$ is defined as:

$$A_j = \frac{P(Test = j, D_j)}{\sum_{k=0}^{K} P(Test = k, D_j)} = \frac{P(Test = j, D_j)}{P(D_j)}, j = 0, \ldots, K.$$

The intrinsic accuracy metric can be written as $A_j = P\big(Test = j \mid D_j\big)$. With a binary disease and test, $A_0$ denotes specificity and $A_1$ denotes sensitivity.

Predictive value *positive* corresponding to test readout category $k = 1, \ldots, K$ is:

$$PVP_k = \frac{P(Test = k, D_k)}{\sum_{j=0}^{J} P(Test = k, D_j)} = \frac{P(Test = k, D_k)}{P(Test = k)}, k = 1, \ldots, K.$$

Predictive value *negative* corresponding to test readout category $k = 0, \ldots, K$ is:

$$PVN_k = \frac{P(Test = k, D_0)}{\sum_{j=0}^{J} P(Test = k, D_j)} = \frac{P(Test = k, D_0)}{P(Test = k)}, k = 0, \ldots, K.$$

The predictive value metrics can be written as $PVP_k = P(D_k \mid Test = k)$ and $PVN_k = P(D_0 \mid Test = k)$. With a binary disease and test, $PVP_1$ is PPV and $PVN_0$ is NPV.

To facilitate contextualization of the methodology developed in this paper, presented in the table directly below is data published for a case-control study of the performance of a blood-based MCED test (4). The generation of this table and detailed analyses are described in Section 7 below. Some important observations are:

i. although the total number of cancer cases are considered fixed by design, the number of cases of each type is random;
ii. for a given cancer type, three types of classifications are possible: Negative, positive with incorrect TOO classification, positive with correct TOO classification;
iii. clinically actionable information necessitates establishing predictive value positive separately for each TOO readout and needs to address the oversampling of cases and the sparse false positive counts in the Control group.

It is clear that, in order to properly interpret such large-dimensional confusion matrices, it is required to formulate focused performance metrics that target key pieces of information that can be of value in aiding clinical decision making.

| | Negative | Uterus | UGI | Prostate | PG | Lung | HN | CRC | Breast | Kidney | Others | TOTAL |
|---|---|---|---|---|---|---|---|---|---|---|---|---|
| **Control** | 606 | 0 | 0 | 0 | 0 | 0 | 0 | 0 | 0 | 0 | 4 | 610 |
| **Uterus** | 27 | 8 | 0 | 0 | 0 | 0 | 0 | 0 | 0 | 0 | 1 | 36 |
| **UGI** | 5 | 0 | 17 | 0 | 1 | 2 | 0 | 1 | 0 | 0 | 0 | 26 |
| **Prostate** | 74 | 0 | 0 | 10 | 0 | 0 | 0 | 0 | 0 | 0 | 0 | 84 |
| **PG** | 6 | 0 | 1 | 0 | 30 | 0 | 0 | 0 | 1 | 0 | 2 | 40 |
| **Lung** | 39 | 0 | 0 | 0 | 0 | 71 | 1 | 0 | 0 | 0 | 0 | 111 |
| **HN** | 8 | 0 | 0 | 0 | 0 | 1 | 15 | 0 | 1 | 0 | 0 | 25 |
| **CRC** | 12 | 0 | 2 | 0 | 0 | 1 | 0 | 38 | 0 | 0 | 0 | 53 |
| **Breast** | 63 | 0 | 0 | 0 | 0 | 0 | 0 | 0 | 40 | 0 | 1 | 104 |
| **Kidney** | 22 | 0 | 0 | 0 | 0 | 0 | 0 | 0 | 0 | 3 | 0 | 25 |
| **Others** | 50 | 0 | 2 | 0 | 0 | 4 | 2 | 0 | 1 | 0 | 91 | 150 |
| **TOTAL** | 912 | 8 | 22 | 10 | 31 | 79 | 18 | 39 | 43 | 3 | 99 | 1264 |

**Figure 6 B. from Liu et al. (2020) (with modification):** modification as described in Section 7 to permit analysis of our proposed suite of performance metrics. UGI: upper gastrointestinal; PG: Pancreas and gallbladder; HN: Head and neck; CRC: colon and rectum.

## 4. Estimation and inference for performance metrics for multi-category tests and disease states using case-control sampling

For a case-control design, both the number of controls, $N_0$, and the number of cases, $N - N_0 = N_1$ are fixed in advance by the design. However, the number of cases of a particular disease state, $n_{j+}, j = 1, \dots, J$, are not in general fixed in advance in the study design; instead, these arise as random variables that are dependent on the underlying disease incidence and study sampling process. In this setting, under random sampling of $N_1$ cases from $\boldsymbol{\mathcal{P}}$, the $\{n_{j+}: j = 1, \dots, J-1\}$ constitute a multinomial random vector with parameters $N_1$ and $\{p_j: j = 1, \dots, J-1\}$, with $p_j = P(D_j \mid D)$ and $p_J = 1 - \sum_{j=1}^{J-1} P(D_j \mid D)$. Maximum likelihood estimates for $P(D_j \mid D)$ are $\hat{p}_j = n_{j+}/N_1$ and $\hat{p}_J = 1 - \sum_{j=1}^{J-1} \hat{p}_j$.

Estimation and inference for predictive value metrics require values for disease incidence $P(D_j)$. Two approaches are considered for this purpose. If a national registry database or published large population study exists for the intended use population, values of $P(D_j)$ can be sourced from it. If the observed data in Table 2 is truly obtained via a conditional random sample of $N_0$ controls and $N_1$ cases from $\boldsymbol{\mathcal{P}}$, because $P(D_j) = P(D_j, D) = P(D_j \mid D)P(D)$, an estimate of $P(D_j)$ may be obtained from multiplying an estimate of $P(D_j \mid D)$ obtained from the case-control dataset and a proposed value for $P(D)$, possibly obtained from an external source or subject matter experts if none exists. Both of these approaches to obtain $P(D_j)$ will be evaluated below.

Define the shorthand notation for test readout category $T_k \equiv (Test = k)$. When $j = 0$, the maximum likelihood estimator for $A_0 = P(T_0 \mid D_0)$ is the Binomial proportion $n_{00}/N_0$. Similarly, the false-positive error corresponding to each test category is defined as $\beta_k = P(T_k \mid D_0), k = 1, \dots, K$, with corresponding estimates $\hat{\beta}_k = n_{0k}/N_0$. Standard confidence intervals for $A_0$ and $\beta_k$ can be obtained, for example the Mid-P approach (5); this approach is used in the Simulation study and Data analysis sections below. In settings with very high specificity, e.g. screening tests, the counts $n_{0k}$ can be sparse when the number of positive test categories *K* is large relative to $N_0$, this can make direct estimation of $P(T_k \mid D_0), k = 1, \dots, K$, challenging. Approaches to address this challenge are proposed in section 4.5 below.

### *4.1 Estimation and inference for intrinsic accuracy metrics*

To derive an estimator for $A_{jk} = P(T_k \mid D_j), j = 1, \dots, J, k = 0, \dots, K$, define the Bernoulli random variable $Y_i^{(jk)} = I(subject\ i\ from\ group\ D_j\ has\ Test = k)$, where $I(a)$ is the indicator function, assuming the value 1 for the event $a$ true, 0 otherwise. Note $Y_i^{(jk)}$ has parameter $P(T_k \mid D_j)$. The sum $n_{jk} = \sum_{i=1}^{n_{j+}} Y_i^{(jk)}$ is a *compound random variable* since both $n_{j+}$ and $Y_i^{(jk)}$ are random variables and $Y_i^{(jk)}$ is considered independent of $n_{j+}$. When the $\{n_{j+}: j = 1, \dots, J-1\}$ arise from multinomial sampling, the random count $n_{j+}$ has a Binomial distribution with parameters $N_1$ and $P(D_j \mid D)$.

Define $\tilde{A}_{jk} = \left(\frac{1}{n_{j+}}\right) \sum_{i=1}^{n_{j+}} Y_i^{(jk)}$. Using the convention that $0/0 \equiv 0$, it can be shown that $E(\tilde{A}_{jk}) = E\left(Y_i^{(jk)}\right) P(n_{j+} > 0)$ and

$$V(\tilde{A}_{jk}) = E^2\left(Y_i^{(jk)}\right) P(n_{j+} > 0)\{1 - P(n_{j+} > 0)\} + V\left(Y_i^{(jk)}\right) E\left(\frac{1}{n_{j+}} \mid n_{j+} > 0\right) P(n_{j+} > 0),$$

where $E(a)$ and $V(a)$ denote expectation and variance of the random variable $a$

with respect to $P(x)$. Consider estimating $A_{jk}$ by $\hat{A}_{jk} = \tilde{A}_{jk}/P(n_{j+} > 0)$. The expectation of $\hat{A}_{jk}$ is:

$$E\left(\frac{\tilde{A}_{jk}}{P(n_{j+} > 0)}\right) = \frac{E(\tilde{A}_{jk})}{P(n_{j+} > 0)} = E\left(Y_i^{(jk)}\right) = P(\,T_k \mid D_j\,)$$

and the variance of $\hat{A}_{jk}$ is

$$\begin{aligned}\sigma_{jk}^2 = V(\hat{A}_{jk}) &= V(\tilde{A}_{jk})/P^2(n_{j+} > 0) \\ &= \frac{1}{P(n_{j+} > 0)}\Bigg[P^2(\,T_k \mid D_j\,)\{1 - P(n_{j+} > 0)\} \\ &+ P(\,T_k \mid D_j\,)\{1 - P(\,T_k \mid D_j\,)\}E\left(\frac{1}{n_{j+}} \mid n_{j+} > 0\right)\Bigg].\end{aligned}$$

The sample analog for $V(\hat{A}_{jk})$, denoted by $\hat{\sigma}_{jk}^2$, is obtained by substituting $\hat{A}_{jk}$ for $P(\,T_k \mid D_j\,)$ in the expression directly above. The quantities $\hat{A}_{jk}$ and $\hat{\sigma}_{jk}^2$ rely on a known distribution for $n_{j+}$. In practice, $\hat{p}_j$ is substituted for $P(\,D_j \mid D\,)$ in the Binomial distribution for $n_{j+}$. Because $N_1$ is typically considered large, errors in this approximation are assumed negligible; this claim is supported in the simulation study.

Consider deriving a confidence interval for $\log[\text{odds}\{P(\,T_k \mid D_j\,)\}] = \text{logit}\{P(\,T_k \mid D_j\,)\}$, which for both analytical and numerical considerations is preferred over direct evaluation of $P(\,T_k \mid D_j\,)$. Write

$$\text{logit}\{P(\,T_k \mid D_j\,)\} = \log\left\{\frac{P(\,T_k \mid D_j\,)}{\sum_{\ell=0}^{K} I(\ell \neq k)P(\,T_\ell \mid D_j\,)}\right\}.$$

An estimator for the above substitutes $\hat{A}_{j\ell}$ for $P(\,T_\ell \mid D_j\,)$, which equals

$$\begin{aligned}\text{L}(\tilde{A}_{jk}) &= \log\left\{\frac{\tilde{A}_{jk}}{\sum_{\ell=0}^{K} I(\ell \neq k)\tilde{A}_{j\ell}}\right\} = \log\left(\frac{n_{jk}/n_{j+}}{\sum_{\ell=0}^{K} I(\ell \neq k)\dfrac{n_{j\ell}}{n_{j+}}}\right) = \log\left(\frac{n_{jk}/n_{j+}}{1 - \dfrac{n_{jk}}{n_{j+}}}\right) \\ &= \log\left(\frac{\tilde{A}_{jk}}{1 - \tilde{A}_{jk}}\right).\end{aligned}$$

The approximate variance of $\text{L}(\tilde{A}_{jk})$ can be obtained using the delta-method, which is evaluated to equal:

$$V\{\mathrm{L}(\tilde{A}_{jk})\} = \frac{P(n_{j+} > 0)\left[P(T_k \mid D_j)\{1 - P(n_{j+} > 0)\} + \{1 - P(T_k \mid D_j)\}E\left(\frac{1}{n_{j+}} \mid n_{j+} > 0\right)\right]}{P(T_k \mid D_j)\{1 - P(T_k \mid D_j)\}^2}.$$

An estimator for $V\{\mathrm{L}(\tilde{A}_{jk})\}$ is obtained by substituting $\hat{A}_{jk}$ for $P(T_k \mid D_j)$ and $\hat{p}_j$ for $P(D_j \mid D)$ in the Binomial distribution for $n_{j+}$, denoted $\hat{V}\{\mathrm{L}(\tilde{A}_{jk})\}$.

A $100(1-\alpha)\%$ Wald-type confidence interval for $\mathrm{logit}\{P(T_k \mid D_j)\}$ is:

$$\mathrm{L}(\tilde{A}_{jk}) \pm z_{1-\alpha/2}\sqrt{\hat{V}\{\mathrm{L}(\tilde{A}_{jk})\}}, \qquad j = 1, \ldots, K.$$

The anti-logit transform of this confidence interval yields the corresponding one for $A_{jk.}$

The intrinsic accuracy estimator of $A_j$ corresponding to disease state $j = 1, \ldots, K$ is $\hat{A}_{jj} = \tilde{A}_{jj}/P(n_{j+} > 0)$ and the estimated logit transform is $\mathrm{L}(\tilde{A}_{jj})$. The metric $1 - A_j$ for $j = 1, \ldots, K$, encompasses two types of misclassification error:

i. cases that obtain a Negative test readout, and
ii. cases that obtain a positive test readout but the incorrect tissue of origin classification.

To enable separation of these two types of misclassification errors, define the metric $A_j^0 = P(T_0 \mid D_j), j = 1, \ldots, J$; this is the false Negative classification error corresponding to disease state $j$. To obtain an estimator and corresponding $100(1-\alpha)\%$ confidence interval for $A_j^0$, the Bernoulli variable $Y_i^{(j0)}$ is used in the inference procedure described above, with ensuing notational changes resulting from setting the value $k = 0$. Note that the value $1 - A_j^0 = 1 - P(T_0 \mid D_j)$ is below referred to as "crude sensitivity" for disease state $j$; it is the probability of not a Negative test and encompasses errors of type ii. defined directly above.

For an overall intrinsic accuracy measurement, the aggregate proportion of subjects across disease states $1, \ldots, K$ that obtain the correct test classification is calculated as:

$$\frac{1}{\sum_{j=1}^{K} n_{j+}} \sum_{j=1}^{K} \sum_{i=1}^{n_{j+}} Y_i^{(jj)} = \frac{1}{\sum_{j=1}^{K} n_{j+}} \sum_{j=1}^{K} n_{jj.}$$

The expected value of this aggregate metric is $E\left\{\frac{1}{\sum_{j=1}^{K} n_{j+}} \sum_{j=1}^{K} n_{j+} P(T_j \mid D_j)\right\}$, where the expectation is with respect to the distribution of $n_{1+}, \ldots, n_{K+}$, so that this aggregate metric does not transcend information for use at a per-subject level.

*4.2 Estimation and inference for predictive value metrics*

For analysis of predictive value metrics, write $P(T_k, D_j) = P(T_k \mid D_j)P(D_j)$ and $P(D_j) = P(D_j, D) = P(D_j \mid D)P(D), j = 1, \ldots, J.$ Because $P(T_k \mid D_j)$ and $P(D_j \mid D)$ are directly estimable with a case-control design and random sampling of cases and controls from $\boldsymbol{\mathcal{P}}$, the general form for $PVP_k$ is:

$$PVP_k = \frac{P(T_k \mid D_k)P(D_k \mid D)P(D)}{P(T_k \mid D_0)\{1 - P(D)\} + P(D)\sum_{j=1}^{J} P(T_k \mid D_j)P(D_j \mid D)}, \quad k = 1, \ldots, K.$$

Similarly,

$$PVN_k = \frac{P(T_k \mid D_0)\{1 - P(D)\}}{P(T_k \mid D_0)\{1 - P(D)\} + P(D)\sum_{j=1}^{J} P(T_k \mid D_j)P(D_j \mid D)}, \quad k = 0, \ldots, K.$$

A value for overall disease incidence $P(D)$ is assumed to be obtained from an external source, such as a population disease registry, published large population study or subject matter experts, and is considered fixed for inferential purposes. Estimators for $PVP_k$ and $PVN_k$ and their variances derived below may be obtained by substitution of $\hat{A}_{jk}$ for $P(T_k \mid D_j)$ and $n_{0k}/N_0$ for $P(T_k \mid D_0)$. Two settings will be considered for obtaining values for $P(D_j \mid D)$, either sourcing from a national registry or plugging in the estimator $\hat{p}_j$.

To conduct inference for the population predictive value, the logit transformation is used along with the delta-method approach for variance formulation. Define the logit transform of $PVP_k$ by:

$$U(\boldsymbol{\varphi_k}) = \log\left\{\frac{P(T_k \mid D_k)P(D_k \mid D)P(D)}{R_U(\boldsymbol{\varphi_k})}\right\}, k = 1, \ldots, K,$$

where

$$R_U(\boldsymbol{\varphi_k}) = P(T_k \mid D_0)\{1 - P(D)\}$$
$$+ P(D)\sum_{j=1}^{J-1} I(j \neq k)\, P(T_k \mid D_j)P(D_j \mid D) + P(D)\, I(J \neq k)P(T_k \mid D_J)\left\{1 - \sum_{j=1}^{J-1} P(D_j \mid D)\right\}.$$

The logit transformation of $PVN_k$ , $k = 0, \ldots, K,$ is:

$$W(\boldsymbol{\varphi_k}) = \log\left\{\frac{P(T_k \mid D_0)\{1 - P(D)\}}{P(D)\sum_{j=1}^{J-1} P(T_k \mid D_j)P(D_j \mid D) + P(D)P(T_k \mid D_J)\left\{1 - \sum_{j=1}^{J-1} P(D_j \mid D)\right\}}\right\}.$$

Here, $\boldsymbol{\varphi_k}$ is the $(2J)$-column vector of parameters:

$$\boldsymbol{\varphi_k} = \{P(T_k \mid D_0), P(T_k \mid D_1), \ldots, P(T_k \mid D_J), P(D_1 \mid D), \ldots, P(D_{J-1} \mid D)\}^t$$

with $\boldsymbol{a}^t$ denoting the transpose of the matrix $\boldsymbol{a}$. Denote the sample analog of $\boldsymbol{\varphi_k}$ by $\widehat{\boldsymbol{\varphi}}_{\boldsymbol{k}}$ that substitutes the corresponding estimators for components of $\boldsymbol{\varphi_k}$ as defined above.

Using the delta-method, the asymptotic variance of the sample version of $\mathrm{logit}(PVP_k)$, denoted $U(\widehat{\boldsymbol{\varphi}}_{\boldsymbol{k}})$, that substitutes $\hat{p}_j$ for $P\big(D_j \mid D\big)$ is:

$$V\{U(\widehat{\boldsymbol{\varphi}}_{\boldsymbol{k}})\} = \{U'(\boldsymbol{\varphi_k})\}^t\, V(\widehat{\boldsymbol{\varphi}}_{\boldsymbol{k}} - \boldsymbol{\varphi_k})\{U'(\boldsymbol{\varphi_k})\}\,,$$

where the column vector $U'(\boldsymbol{\varphi_k})$ is:

$$\{U'(\boldsymbol{\varphi_k})\}^t = \left(\frac{\partial U(\boldsymbol{\varphi_k})}{\partial P(T_k \mid D_0)}, \frac{\partial U(\boldsymbol{\varphi_k})}{\partial P(T_k \mid D_1)}, \dots, \frac{\partial U(\boldsymbol{\varphi_k})}{\partial P(T_k \mid D_J)}, \frac{\partial U(\boldsymbol{\varphi_k})}{\partial P(D_1 \mid D)}, \cdots, \frac{\partial U(\boldsymbol{\varphi_k})}{\partial P(D_{J-1} \mid D)}\right)$$

and $V(\widehat{\boldsymbol{\varphi}}_{\boldsymbol{k}} - \boldsymbol{\varphi_k})$ is the $(2J \times 2J)$ variance-covariance matrix of $\widehat{\boldsymbol{\varphi}}_{\boldsymbol{k}}$. The components of $\{U'(\boldsymbol{\varphi_k})\}^t$ and $V(\widehat{\boldsymbol{\varphi}}_{\boldsymbol{k}} - \boldsymbol{\varphi_k})$ are provided in the Appendix.

When the terms $P\big(D_j \mid D\big)$ are considered fixed, for example when obtained from a population registry, the asymptotic variance of the corresponding version of $U(\widehat{\boldsymbol{\varphi}}_{\boldsymbol{k}})$ is

$$V_0\{U(\widehat{\boldsymbol{\varphi}}_{\boldsymbol{k}})\} = \{U_0'(\boldsymbol{\varphi_k})\}^t\, V_0(\widehat{\boldsymbol{\varphi}}_{\boldsymbol{k}} - \boldsymbol{\varphi_k})\{U_0'(\boldsymbol{\varphi_k})\}\,,$$

where $U_0'$ and $V_0$ are as $U'$ and $V$ except with all components involving $P\big(D_j \mid D\big)$ set to equal 0.

Similarly, using the delta-method, the asymptotic variance of the sample version of logit($PVN_k$), denoted $W(\widehat{\boldsymbol{\varphi}}_{\boldsymbol{k}})$, that substitutes $\hat{p}_j$ for $P\big(D_j \mid D\big)$ is:

$$V\{W(\widehat{\boldsymbol{\varphi}}_{\boldsymbol{k}})\} = \{W'(\boldsymbol{\varphi_k})\}^t\, V(\widehat{\boldsymbol{\varphi}}_{\boldsymbol{k}} - \boldsymbol{\varphi_k})\{W'(\boldsymbol{\varphi_k})\}\,,$$

where the column vector $W'(\boldsymbol{\varphi_k})$ is:

$$\{W'(\boldsymbol{\varphi_k})\}^t = \left(\frac{\partial W(\boldsymbol{\varphi_k})}{\partial P(T_k \mid D_0)}, \frac{\partial W(\boldsymbol{\varphi_k})}{\partial P(T_k \mid D_1)}, \dots, \frac{\partial W(\boldsymbol{\varphi_k})}{\partial P(T_k \mid D_J)}, \frac{\partial W(\boldsymbol{\varphi_k})}{\partial P(D_1 \mid D)}, \cdots, \frac{\partial W(\boldsymbol{\varphi_k})}{\partial P(D_{J-1} \mid D)}\right).$$

The components of $\{W'(\boldsymbol{\varphi_k})\}^t$are provided in the Appendix. When the $P\big(D_j \mid D\big)$ are fixed, the asymptotic variance of $W(\widehat{\boldsymbol{\varphi}}_{\boldsymbol{k}})$ takes the form

$$V_0\{W(\widehat{\boldsymbol{\varphi}}_{\boldsymbol{k}})\} = \{W_0'(\boldsymbol{\varphi_k})\}^t\, V_0(\widehat{\boldsymbol{\varphi}}_{\boldsymbol{k}} - \boldsymbol{\varphi_k})\{W_0'(\boldsymbol{\varphi_k})\}\,,$$

Where $W_0'$ is as $W'$ except with all components involving $P\big(D_j \mid D\big)$ set to equal 0.

When estimating $P\big(D_j \mid D\big)$ with $\hat{p}_j$, a $100(1-\alpha)\%$ Wald-type confidence interval for $\mathrm{logit}\{PVP_k\}$ is

$$U(\widehat{\boldsymbol{\varphi}}_{\boldsymbol{k}}) \pm z_{1-\alpha/2}\sqrt{\widehat{V}\{U(\widehat{\boldsymbol{\varphi}}_{\boldsymbol{k}})\}}, k = 1, \dots, K,$$

where $\hat{V}\{U(\hat{\boldsymbol{\varphi}}_{\boldsymbol{k}})\}$ arises from substituting sample analogs for population parameters in $V\{U(\hat{\boldsymbol{\varphi}}_{\boldsymbol{k}})\}$; when $P\big(D_j \mid D\big)$ is considered fixed, $\hat{V}_0\{U(\hat{\boldsymbol{\varphi}}_{\boldsymbol{k}})\}$ replaces $\hat{V}\{U(\hat{\boldsymbol{\varphi}}_{\boldsymbol{k}})\}$. A $100(1-\alpha)\%$ Wald-type confidence interval for $\mathrm{logit}\{PVN_k\}$ can analogously be constructed. The anti-logit transform of the corresponding confidence interval yields the one for $PVP_k$ or $PVN_k$.

### *4.3 Estimation and inference for marginal test category distribution*

Obtaining estimation and inference procedures for the marginal probabilities $P(T_k) = P(Test = k), k = 0, \ldots, K$, provides useful information for interpreting the test's potential impact in the population and for designing prospective cohort studies. Inference for $P(T_k)$ can greatly be simplified if the full distribution in Table 2 is collapsed as:

| | ***Test=0*** | ***Test=1*** | $\cdots$ | ***Test=K*** | **TOTAL** |
|---|---|---|---|---|---|
| **Controls**$(\boldsymbol{D_0})$ | $n_{00}$ | $n_{01}$ | $\cdots$ | $n_{0K}$ | $N_0$ |
| **Cases** $(\boldsymbol{D})$ | $n_{+0} - n_{00}$ | $n_{+1} - n_{01}$ | $\cdots$ | $n_{+K} - n_{0K}$ | $N_1$ |
| **TOTAL** | $n_{+0}$ | $n_{+1}$ | $\cdots$ | $n_{+K}$ | $N$ |

**Table 3**: Collapsed version of Table 2 to facilitate inference for marginal Test category distribution.

Write $P(T_k) = P(T_k \mid D_0)\{1 - P(D)\} + P(T_k \mid D)P(D), k = 0, \ldots, K.$ The logit transformation of $P(T_k)$ is:

$$Q(\boldsymbol{\vartheta}_k) = \log\left[\frac{P(T_k \mid D_0)\{1 - P(D)\} + P(T_k \mid D)P(D)}{\{1 - P(T_k \mid D_0)\}\{1 - P(D)\} + \{1 - P(T_k \mid D)\}P(D)}\right] = \log\left(\frac{B_k^0}{B_k^1}\right),$$

where $\boldsymbol{\vartheta}_k^t = \big(P(T_k \mid D_0), P(T_k \mid D)\big)$. The sample analog of $\boldsymbol{\vartheta_k}$ is denoted by $\hat{\boldsymbol{\vartheta}}_{\boldsymbol{k}}$, that substitutes the Binomial proportions $n_{0k}/N_0$ for $P(T_k \mid D_0)$ and $(n_{+k} - n_{0k})/N_1$ for $P(T_k \mid D)$. Using the delta-method, the asymptotic variance of $Q(\hat{\boldsymbol{\vartheta}}_{\boldsymbol{k}})$, the sample version of logit$\{P(T_k)\}$, is $V\{Q(\hat{\boldsymbol{\vartheta}}_{\boldsymbol{k}})\} = \{Q'(\boldsymbol{\vartheta_k})\}^t\, V(\hat{\boldsymbol{\vartheta}}_{\boldsymbol{k}} - \boldsymbol{\vartheta_k})\{Q'(\boldsymbol{\vartheta_k})\}$,

where the column vector $Q'(\boldsymbol{\vartheta_k})$ is:

$$\{Q'(\boldsymbol{\vartheta_k})\}^t = \left(\frac{1 - P(D)}{B_k^0 B_k^1}, \frac{P(D)}{B_k^0 B_k^1}\right)$$

and $V(\hat{\boldsymbol{\vartheta}}_{\boldsymbol{k}} - \boldsymbol{\vartheta_k})$ is the $(2 \times 2)$ diagonal matrix, with diagonal equal to

$$[N_0^{-1} P(T_k \mid D_0)\{1 - P(T_k \mid D_0)\},\ N_1^{-1} P(T_k \mid D)\{1 - P(T_k \mid D)\}].$$

A $100(1-\alpha)\%$ Wald-type confidence interval for $\mathrm{logit}\{P(T_k)\}$ is:

$$Q(\hat{\boldsymbol{\vartheta}}_{\boldsymbol{k}}) \pm z_{1-\alpha/2}\sqrt{\hat{V}\{Q(\hat{\boldsymbol{\vartheta}}_{\boldsymbol{k}})\}}, \qquad k = 1, \ldots, K,$$

where $\hat{V}\{Q(\widehat{\boldsymbol{\vartheta}}_{\boldsymbol{k}})\}$ is obtained from substituting sample analogs for population parameters in $V\{Q(\widehat{\boldsymbol{\vartheta}}_{\boldsymbol{k}})\}$; the anti-logit transform yields the one for $P(T_k)$.

### *4.4 Estimation and inference for an overall predictive value metric*

An *overall* Predictive value *positive* metric corresponding to all positive Test readout categories $k = 1, \dots, K$ is defined as:

$$PVP^* = \frac{\sum_{k=1}^{K} P(T_k, D_k)}{\sum_{k=1}^{K} P(T_k)} = \sum_{k=1}^{K} P(D_k \mid T_k) \left\{ \frac{P(T_k)}{\sum_{\ell=1}^{K} P(T_k)} \right\}$$
$$= \frac{\sum_{k=1}^{K} P(T_k \mid D_k) P(D_k \mid D) P(D)}{\sum_{k=1}^{K} \left[ \sum_{j=1}^{J} P(T_k \mid D_j) P(D_j \mid D) P(D) + P(T_k \mid D_0)\{1 - P(D)\} \right]}.$$

As was the situation when considering an aggregate intrinsic accuracy metric, an aggregate predictive value metric has no clear population analog to facilitate its interpretation at a per-subject level, and its development is not further considered.

### *4.5 Adjustments for sparse false-positive counts*

In settings that require a very high specificity target, for example in a test intended for general population screening, the number of false-positive counts $n_{0k}$ (i.e., controls misclassified into positive test category $T_k, k = 1, \dots, K$) can be sparse, especially when the number of positive readout categories $K$ is large relative to $N_0$. In these settings, because of the large weight $1 - P(D)$ assigned to $P(T_k \mid D_0)$ in the analytical form of $PVP_k$, $PVN_k$ and the marginal test distribution, this sparseness can lead to instability in estimation and inference procedures for these metrics. Of particular concern are settings with $n_{0k} = 0$, referred to as "sampling zeros", that result in an estimate of $P(T_k \mid D_0)$ that values 0.

Recall the sampling model for the control group test readout counts is multinomial with parameters $N_0$ and $\{P(T_k \mid D_0): k = 0, \dots, K\}$, which is a saturated model. Authors have advocated for adding ½ to each cell count for saturated models with sparse data in order to reduce bias in resulting estimation (6,7,8). Using this approach, we obtain $n_{0k}^* = n_{0k} + \frac{1}{2}, k = 0, \dots, K$. The resulting estimators are $\tilde{P}(T_k \mid D_0) = n_{0k}^*/\{N_0 + 0.5(K+1)\}$. The variance of the estimator $\tilde{P}(T_k \mid D_0)$ is obtained by multiplying the multinomial variance by $N_0^2/\{N_0 + 0.5(K+1)\}^2$, which approaches 1 as $N_0$ increases so that no adjustments are explicitly made to variance formulas when implementing this adjustment. This adjustment may also be warranted for disease states when estimating $P(T_k \mid D_j)$ with sparse observed data and relatively high conditional incidence $P(D_j \mid D), j = 1, \dots, J$.

## 5. Stratified analyses using case-control sampling

Analyses of subgroups of Table 2 arising from partitions defined by factors other than disease or test readout variables, e.g. demographic factors, proceed naturally using the corresponding methods above on the partitioned data. Stratified analysis of predictive-value and marginal test distribution metrics require stratum-specific disease incidence values that may not be available from an external source; in such settings, conditional incidence estimates given disease derived from the observed data under a random sampling model can be paired with an overall incidence value.

When it is of interest to stratify analysis based on disease stage, two situations arise. Intrinsic accuracy types of analyses proceed directly using the methodology above, where a conditional test readout distribution corresponding to a particular disease stage is analyzed. In contrast, when investigating predictive value positive for a given test readout category, the (overall) metric can be decomposed into sub-parts corresponding to each disease stage. Let the integer valued variable $S$ denote disease stage. Define

$$PVP_k^{(S)} = \frac{P(T_k \mid D_k, D, S)P(S \mid D_k, D)P(D_k \mid D)P(D)}{P(T_k \mid D_0)\{1 - P(D)\} + P(D)\sum_{j=1}^{J} P(T_k \mid D_j)P(D_j \mid D)}, \quad k = 1, \ldots, K.$$

An estimator of $PVP_k^{(S)}$ is obtained analogously as for $PVP_k$ with either the sample-based analog estimator for $P(\,S \mid D_k, D\,)$ or one sourced from a national registry or published large population study.

To conduct inference for $PVP_k^{(S)}$ at the stage value $S = s_0$, the logit transform is:

$$Z_{S=s_0}(\boldsymbol{\psi_k}) = \log\left\{\frac{P(T_k \mid D_k, S = s_0)P(S = s_0 \mid D_k)P(D_k \mid D)P(D)}{R_U(\boldsymbol{\varphi_k}) + P(T_k \mid D_k, S \neq s_0)P(S \neq s_0 \mid D_k)P(D_k \mid D)P(D)}\right\}, k = 1, \ldots, K,$$

where

$$\boldsymbol{\psi_k} = \{\boldsymbol{\varphi_k}, P(T_k \mid D_k, S = s_0), P(T_k \mid D_k, S \neq s_0), P(S = s_0 \mid D_k), P(S \neq s_0 \mid D_k)\}.$$

The delta-method may be used as shown above to obtain a variance estimator for $Z_{S=s_0}(\widehat{\boldsymbol{\psi}}_{\boldsymbol{k}})$, where $\widehat{\boldsymbol{\psi}}_{\boldsymbol{k}}$ is the sample analog of $\boldsymbol{\psi_k}$, from which an asymptotic $100(1-\alpha)\%$ Wald-type confidence interval for $\text{logit}\left\{PVP_k^{(S)}\right\}$ can analogously be constructed.

## 6. Simulation Study

A numerical study was conducted to investigate the performance of the proposed methodology in several simulated real-world scenarios. Two general settings were considered, corresponding to either a screening test or diagnostic test. The general format used *J=3* disease states and *K=2* positive test readouts as displayed below.

| | ***Test=0*** $(T_0)$ | ***Test=1*** $(T_1)$ | ***Test=2*** $(T_2)$ | **TOTAL** |
|---|---|---|---|---|
| $D_0$ | $E(n_{00}) = N_0 SP_0$ | $E(n_{01}) = N_0 \beta_1$ | $n_{02} = N_0 - n_{00} - n_{01}$ | $N_0$ |
| $D_1$ | $E(n_{10}) = n_{1+}\alpha_1$ | $E(n_{11}) = n_{1+}SE_1$ | $n_{12} = n_{1+} - n_{10} - n_{11}$ | $E(n_{1+}) = N_1 p_1$ |
| $D_2$ | $E(n_{20}) = n_{2+}\alpha_2$ | $n_{21} = n_{2+} - n_{20} - n_{22}$ | $E(n_{22}) = n_{2+}SE_2$ | $E(n_{2+}) = N_1 p_2$ |
| $D_3$ | $E(n_{30}) = n_{3+}\alpha_{31}$ | $E(n_{31}) = n_{3+}\alpha_{32}$ | $n_{32} = n_{3+} - n_{30} - n_{31}$ | $n_{3+} = N_1 - n_{1+} - n_{2+}$ |
| **TOTAL** | $n_{+0}$ | $n_{+1}$ | $n_{+2}$ | $N = N_0 + N_1$ |

**Table 4:** Two-way frequency distribution of disease and test for simulation study.

The general sampling framework is described as follows. The number of controls, $N_0$ and the total number of cases, $N_1$, are fixed in advance. The number of enrolled cases of each disease state follows a multinomial sampling distribution with parameters $N_1$ and $(p_1, p_2, p_3 = 1 - p_1 - p_2)$. The value of the test is ascertained on each subject according to the following metrics:

| **Notation** | $SP_0$ | $\beta_1$ | $\alpha_1$ | $A_1$ | $\alpha_2$ | $A_2$ | $\alpha_{31}$ | $\alpha_{32}$ |
|---|---|---|---|---|---|---|---|---|
| **Metric** | $P(T_0 \mid D_0)$ | $P(T_1 \mid D_0)$ | $P(T_0 \mid D_1)$ | $P(T_1 \mid D_1)$ | $P(T_0 \mid D_2)$ | $P(T_2 \mid D_2)$ | $P(T_0 \mid D_3)$ | $P(T_1 \mid D_3)$ |
| **Screen** | 0.98 | 0.01 | 0.25 | 0.65 | 0.30 | 0.50 | 0.60 | 0.20 |
| **Diagnostic** | 0.50 | 0.25 | 0.03 | 0.95 | 0.05 | 0.92 | 0.10 | 0.40 |

**Table 5:** Conditional test distribution given disease state for screening, diagnostic settings.

Given a choice for population disease incidence, $P(D)$, values for predictive value positive, predictive value negative and the marginal test distribution corresponding to a setting of Table 5 can be directly calculated.

First, for a cancer screening test setting, the presumed goal is to maximize cancer detection while maintaining high specificity in order to minimize the potentially high-risk burden associated with downstream clinical workup of cancer-free individuals. The intended-use population is considered a high-risk population that would benefit from screening, so that $P(D) = 0.016$. For the cancer diagnostic setting, the goal is to minimize false negative results in an intended use population with suspicion of cancer, for example presenting with clinical signs, symptoms and/or findings. Here, the cancer incidence is taken as $P(D) = 0.07$. For both settings, two choices for the distribution of case states among all $N_1$ cases were considered: $p_1 = P(D_1 \mid D) = 0.5$, $p_2 = P(D_2 \mid D) = 0.4$, and, $p_1 = P(D_1 \mid D) = 0.5$, $p_2 = P(D_2 \mid D) = 0.1$. Each simulation exercise generated 10,000 independent realizations from the given population setting.

Tables 6a, 6b and 6c present the mean of bias, the 95% confidence interval (CI) coverage indicator and CI width for target parameters across simulation iterations in the screening setting under several sample sizes for the particular scenario $p_1 = P(D_1 \mid D) = 0.5$, $p_2 = P(D_2 \mid D) = 0.4$.

| | Value | Bias | Coverage | Width |
|---|---|---|---|---|
| $SP_0$ | 98 | 0 | 94.60 | 2.438 |
| $A_1$ | 65 | 0.006 | 95.35 | 11.774 |
| $A_2$ | 50 | -0.023 | 95.30 | 13.772 |
| $PVN_0$ | 99.50 | 0 | 95.04 | 0.134 |
| $PVP_1$ | 31.25 | 2.35 | 93.32 | 32.794 |
| $PVP_2$ | 22.60 | 2.531 | 94.36 | 30.072 |
| $P(T_0)$ | 96.92 | 0 | 94.68 | 2.438 |
| $P(T_1)$ | 1.66 | 0 | 93.44 | 1.74 |
| $P(T_2)$ | 1.42 | 0 | 93.08 | 1.766 |

**Table 6a:** Screening simulation results for $N_0 = N_1 = 500$; all numbers are percents.

| | Value | Bias | Coverage | Width |
|---|---|---|---|---|
| $SP_0$ | 98 | 0.002 | 93.70 | 1.695 |
| $A_1$ | 65 | -0.035 | 95.32 | 8.347 |
| $A_2$ | 50 | -0.002 | 94.92 | 9.767 |
| $PVN_0$ | 99.50 | 0 | 94.71 | 0.094 |
| $PVP_1$ | 31.25 | 1.15 | 94.54 | 23.448 |
| $PVP_2$ | 22.60 | 1.209 | 94.89 | 20.939 |
| $P(T_0)$ | 96.92 | 0.003 | 93.74 | 1.717 |
| $P(T_1)$ | 1.66 | -0.001 | 94.22 | 1.225 |
| $P(T_2)$ | 1.42 | -0.001 | 94.37 | 1.234 |

**Table 6b:** Screening simulation results for $N_0 = N_1 = 1000$; all numbers are percents.

| | Value | Bias | Coverage | Width |
|---|---|---|---|---|
| $SP_0$ | 98 | 0.003 | 94.49 | 1.18 |
| $A_1$ | 65 | 0.002 | 95.12 | 5.906 |
| $A_2$ | 50 | -0.019 | 94.76 | 6.918 |
| $PVN_0$ | 99.50 | 0 | 95.09 | 0.067 |
| $PVP_1$ | 31.25 | 0.513 | 94.61 | 16.494 |
| $PVP_2$ | 22.60 | 0.629 | 95.23 | 14.506 |
| $P(T_0)$ | 96.92 | 0.003 | 94.77 | 1.211 |
| $P(T_1)$ | 1.66 | 0.002 | 94.31 | 0.865 |
| $P(T_2)$ | 1.42 | -0.005 | 94.62 | 0.865 |

**Table 6c:** Screening simulation results for $N_0 = N_1 = 2000$; all numbers are percents.

Tables 7a, 7b and 7c present the mean of bias, the 95% CI coverage indicator and CI width for target parameters in the diagnostic setting under several sample sizes for the scenario $p_1 = P(D_1 \mid D) = 0.5$, $p_2 = P(D_2 \mid D) = 0.4$.

| | Value | Bias | Coverage | Width |
|---|---|---|---|---|
| $SP_0$ | 50 | 0.005 | 93.94 | 8.689 |
| $A_1$ | 95 | 0 | 95.79 | 5.572 |
| $A_2$ | 92 | 0.021 | 95.55 | 7.657 |
| $PVN_0$ | 99.33 | 0.001 | 95.08 | 0.567 |
| $PVP_1$ | 12.34 | 0.076 | 95.06 | 3.859 |
| $PVP_2$ | 9.82 | 0.033 | 95.42 | 3.375 |
| $P(T_0)$ | 46.81 | 0.003 | 94.36 | 8.129 |
| $P(T_1)$ | 26.94 | -0.032 | 95.11 | 7.066 |
| $P(T_2)$ | 26.25 | 0.029 | 94.88 | 7.073 |

**Table 7a:** Diagnostic simulation results for $N_0 = N_1 = 500$; all numbers are percents.

| | Value | Bias | Coverage | Width |
|---|---|---|---|---|
| $SP_0$ | 50 | 0.011 | 94.83 | 6.143 |
| $A_1$ | 95 | 0.009 | 94.98 | 3.875 |
| $A_2$ | 92 | -0.019 | 95.13 | 5.373 |
| $PVN_0$ | 99.33 | 0 | 94.91 | 0.396 |
| $PVP_1$ | 12.34 | 0.034 | 95.15 | 2.713 |
| $PVP_2$ | 9.82 | 0.019 | 94.83 | 2.378 |
| $P(T_0)$ | 46.81 | 0.01 | 95.11 | 5.758 |
| $P(T_1)$ | 26.94 | -0.01 | 94.97 | 5.004 |
| $P(T_2)$ | 26.25 | 0 | 94.80 | 5.005 |

**Table 7b:** Diagnostic simulation results for $N_0 = N_1 = 1000$; all numbers are percents.

| | Value | Bias | Coverage | Width |
|---|---|---|---|---|
| $SP_0$ | 50 | -0.017 | 94.86 | 4.33 |
| $A_1$ | 95 | -0.01 | 95.23 | 2.725 |
| $A_2$ | 92 | 0.007 | 94.91 | 3.775 |
| $PVN_0$ | 99.33 | -0.001 | 95.20 | 0.278 |
| $PVP_1$ | 12.34 | 0.012 | 95.11 | 1.913 |
| $PVP_2$ | 9.82 | 0.009 | 95.10 | 1.676 |
| $P(T_0)$ | 46.81 | -0.016 | 94.98 | 4.075 |
| $P(T_1)$ | 26.94 | -0.003 | 94.98 | 3.541 |
| $P(T_2)$ | 26.25 | 0.019 | 95.19 | 3.542 |

**Table 7c:** Diagnostic simulation results for $N_0 = N_1 = 2000$; all numbers are percents.

For both settings considered, coverage of the 95% CIs was sustained for all scenarios. Bias was very low for the diagnostic setting where the likelihood of a test positive readout was relatively large (about 27% for each positive test category). For the screening settings, the estimators of positive predictive value are based on very small fractions of the dataset (about 1.5% for each positive readout); the bias of these estimators decreased from about 2% to 1% to 0.5% as sample size increased. As expected, the widths of all CIs became narrower with increasing sample size. Because the predictive value positive metrics were much smaller in the diagnostic setting, the CI widths were roughly 10-fold shorter compared to the screening setting.

When comparing results between scenarios with $p_2 = 0.4$ and $p_2 = 0.1$, because the value for $PVP_2$ is smaller when $p_2 = 0.1$, in turn, CI widths also decreased. For example, in the screening setting, for *N=500,* the CI width shrunk from about 30% with $p_2 = 0.4$ to about 12% with $p_2 = 0.1$.

The impact of considering cancer incidence rates fixed, for example when obtained from a population registry, was generally small, with the largest effect corresponding to smaller magnitudes of predictive value positive (less than 15%), with the width of corresponding CIs being roughly 0.5%, 0.4% and 0.3% shorter for *N=500, 1000, 2000*, respectively for both screening and diagnostic settings.

Finally, a numerical study was conducted in a setting where specificity was very high so that an investigation of the proposed adjustment to the control group classification counts was warranted. The screening test setting described above was evaluated where specificity was increased from 0.98 to 0.995, with $p_1 = P(D_1 \mid D) = 0.5$, $p_2 = P(D_2 \mid D) = 0.1$. Table 8 shows results for predictive value positive and marginal test distribution inference for three sample sizes. The expected false-positive count for each positive test category was 1.25, 2.5 and 5 for sample sizes 500, 1000 and 2000, respectively. Confidence interval coverage corresponding to inference on unadjusted data was observed to be very low for $N_0 = N_1 = 500$, however did improve with increasing sample size. Similarly, bias of predictive value positive estimators on unadjusted data was large for $N_0 = N_1 = 500$ but did become smaller as sample size increased; estimators for the marginal test distribution appeared unbiased. Inference corresponding to adjusted false-positive counts enjoyed great improvement in confidence interval coverage compared to the unadjusted counterparts, especially for the smallest sample size configuration. Furthermore, estimation with adjusted false-positive counts greatly reduced bias for $PVP_2$ compared to unadjusted data; bias for $PVP_1$ was similar between adjusted and unadjusted data. Analysis of both adjusted and unadjusted data appeared unbiased for the marginal test distribution.

Based on this simulation result, a general rule-of-thumb for implementing the adjustment to control group classification counts suggests its use in settings where the expected false-positive count for one or more positive test readout categories is less than 5 subjects.

|  | Value | Bias | Coverage | Width |
|---|---|---|---|---|
| $N_0 = N_1 = 500$ |  |  |  |  |
| $\boldsymbol{PVP_1}$ | 56.156 | -3.023 (2.834) | 95.51 (68.98) | 44.75 (36.26) |
| $\boldsymbol{PVP_2}$ | 14.981 | -0.813 (2.563) | 95.84 (73.23) | 24.187 (22.46) |
| $\boldsymbol{P(T_0)}$ | 98.54 | -0.191 (-0.004) | 95.94 (90.81) | 1.44 (1.18) |
| $\boldsymbol{P(T_1)}$ | 0.926 | 0.093 (0.002) | 97.61 (70.77) | 1.006 (0.78) |
| $\boldsymbol{P(T_2)}$ | 0.534 | 0.098 (0.002) | 96.13 (69.45) | 1.066 (0.82) |
| $N_0 = N_1 = 1000$ |  |  |  |  |
| $\boldsymbol{PVP_1}$ | 56.156 | -1.576 (1.316) | 92.74 (89.64) | 33.65 (32.139) |
| $\boldsymbol{PVP_2}$ | 14.981 | -0.283 (1.352) | 93.4 (89.21) | 17.635(18.075) |
| $\boldsymbol{P(T_0)}$ | 98.54 | -0.098 (-0.002) | 95.4 (92.66) | 0.942 (0.855) |
| $\boldsymbol{P(T_1)}$ | 0.926 | 0.049 (0.004) | 90.7 (90.93) | 0.662 (0.594) |
| $\boldsymbol{P(T_2)}$ | 0.534 | 0.049 (-0.002) | 89.51 (90.09) | 0.678 (0.601) |
| $N_0 = N_1 = 2000$ |  |  |  |  |
| $\boldsymbol{PVP_1}$ | 56.156 | -0.647 (0.907) | 93.43 (93.50) | 24.99 (24.846) |
| $\boldsymbol{PVP_2}$ | 14.981 | -0.107 (0.680) | 94.44 (92.87) | 12.82 (13.261) |
| $\boldsymbol{P(T_0)}$ | 98.54 | -0.047 (0.004) | 94.69 (93.09) | 0.638 (0.606) |
| $\boldsymbol{P(T_1)}$ | 0.926 | 0.023 (-0.002) | 94.40 (91.60) | 0.45 (0.427) |
| $\boldsymbol{P(T_2)}$ | 0.534 | 0.024 (-0.002) | 93.29 (93.51) | 0.456 (0.432) |

**Table 8:** Performance with adjusted control group classification counts in screening setting with $SP_0 = 0.995$ and various sample sizes; results for original (unadjusted) data shown in parentheses; all numbers are percents.

## 7. Data Analysis Example

The proposed MCED analysis framework was applied to a published real-world dataset (4). Here, the performance of a multi-cancer detection test using methylation signatures from cell-free DNA that provides tissue localization for positive tests was evaluated using a case-control design. The results reported for the Validation dataset is the focus of analysis here. This dataset consisted of $N_0 = 610$ non-cancer controls and $N_1 = 654$ cancer cases. Cancer incidence rates were obtained using the 2022 SEER database (9). The results (4) show "crude sensitivity" performance of this test, disregarding incorrect TOO classification, by cancer stage in their Figure 5, and predictions of TOO for multiple types of cancer cases in their Figure 6. Because the implementation of the full suite of proposed methods described in our paper requires data on test negative classifications for each cancer type, the analysis reported herein restricts attention to the cancer types reported in Figure 5 (4). Because the data reported in Figure 6 B. (4) did not directly permit construction of a two-way frequency distribution in the format of Table 2, the following two assumptions were necessary. There were 4 stomach cancer cases that received TOO, but the total number of such cases were unreported; we assumed there were 5 total stomach cancer cases with 1 false-negative prediction since overall crude sensitivity (disregarding TOO) was reported as 76.4% (p. 753, second sentence). There was 1 gallbladder case receiving TOO, the total number of such cases were not reported; we assumed there was 1 gallbladder case and no false-negative predictions.

Note that, although Figure 5 (4) presents crude sensitivity by cancer type and stage, this metric only evaluated whether the subject was test “positive” or negative, without describing the TOO prediction accuracy per case type. That is, case types with a positive but incorrect TOO prediction were not considered an error in the crude sensitivity calculation in Figure 5 (4). Using our proposed methods, Table 9 below shows cancer-specific false-negative and intrinsic accuracy estimates by cancer type for all stages. Note that, unlike the results reported in Figure 5 (4), the results in Table 9 below consider a positive test with incorrect TOO as a classification error, which is more clinically relevant. Because the Figure 6 B. (4) is not stratified on cancer stage we were not able to conduct a stage-stratified analysis.

| | **False Negative** | **Intrinsic Accuracy IA(k)** | **IA(k): CI-low** | **IA(k): CI-up** |
|---|---|---|---|---|
| **Uterus** | 75.0 | 22.2 | 11.4 | 38.8 |
| **Upper GI** | 19.2 | 65.4 | 45.3 | 81.2 |
| **Prostate** | 88.1 | 11.9 | 6.50 | 20.8 |
| **Pancreas & Gallbladder** | 15.0 | 75.0 | 59.2 | 86.1 |
| **Lung** | 35.1 | 64.0 | 54.6 | 72.4 |
| **Head & Neck** | 32.0 | 60.0 | 39.9 | 77.2 |
| **CRC** | 22.6 | 71.7 | 58.1 | 82.2 |
| **Breast** | 60.6 | 38.5 | 29.6 | 48.2 |
| **Kidney** | 88.0 | 12.0 | 3.80 | 31.8 |
| **Others** | 33.3 | 60.7 | 52.6 | 68.2 |

**Table 9:** Empirical false-negative proportion and intrinsic accuracy with 95% confidence interval (CI) for 9 cancer types and “Others”. All numbers are percents. CRC: colon and rectum; GI: gastrointestinal.

In consideration of “TOO accuracy”, an “overall accuracy” of 93% was reported (4). This metric is not useful for assessing clinical utility of this test for at least two main reasons. First, this is an aggregate calculation, not providing information per-TOO readout category. Second, the calculation conditions on those cancer case types targeted by the test that had a positive (not Negative) readout; prospectively, a subject’s cancer status and type is unknown at test issuance. Actionable information for a practicing physician from such an MCED test will quantify in the intended use population the likelihood that a given TOO readout corresponds to a cancer case in the corresponding bodily site; this is measured by TOO-specific predictive value positive. Estimates and confidence intervals for predictive value positive and the marginal test distribution for the data reported in Figure 6 B. (4) are provided below in Table 10. Because of the sparse false positive counts, the proposed correction that adds ½ to each non-cancer classification count was implemented.

| | P(T=k) | P(T=k): CI-low | P(T=k): CI-up | PVP(k) | PVP(k): CI-low | PVP(k): CI-up |
|---|---|---|---|---|---|---|
| **Non-Cancer** | 97.85 | 96.68 | 98.61 | 99.4* | 99.3* | 99.4* |
| **Uterus** | 0.10 | 0.01 | 0.96 | 10.4 | 0.6 | 67.6 |
| **Upper GI** | 0.12 | 0.02 | 0.74 | 20.4 | 2.1 | 74.9 |
| **Prostate** | 0.10 | 0.01 | 0.91 | 23.4 | 1.8 | 83.7 |
| **Pancreas & Gallbladder** | 0.14 | 0.03 | 0.68 | 31.3 | 2.9 | 87.5 |
| **Lung** | 0.24 | 0.09 | 0.61 | 52.6 | 10.4 | 91.4 |
| **Head & Neck** | 0.12 | 0.02 | 0.78 | 23.1 | 2.3 | 79.6 |
| **CRC** | 0.16 | 0.04 | 0.65 | 49.1 | 5.9 | 93.7 |
| **Breast** | 0.17 | 0.04 | 0.63 | 43.6 | 5.5 | 91.2 |
| **Kidney** | 0.09 | 0.01 | 1.12 | 7 | 0.4 | 59.7 |
| **Others** | 0.91 | 0.44 | 1.89 | 27.1 | 12.9 | 48.3 |

**Table 10:** Estimated marginal test readout distribution and predictive value positive with 95% confidence interval (CI) for 9 cancer types and “Others”. Predictive value negative indicated with an “*”. All numbers are percents. CRC: colon and rectum; GI: gastrointestinal.

Regarding the marginal test distribution, a negative result is expected in 97.85% of all individuals from the target population receiving the test. Of the 2.15% of individuals receiving a positive test, 0.91% are expected to have an “Other” TOO readout; note this is more than 42% of all positive test readouts. Predictive value negative was very high at 99.4%, as expected. Values for predictive value positive ranged in magnitude from 7% for kidney to 52.6% for Lung, however, all 95% confidence intervals were very wide, resulting in much uncertainty in the clinical utility value of this test base on this dataset.

A graphical illustration that attempts to summarize the “cost-benefit” of an MCED test is shown below in Figure 1 for the test (4). Here, a “target region” is outlined as having a “benefit”, that is, intrinsic accuracy for cancer detection, of at least 0.5 and a “cost”, that is, incorrect TOO prediction, not more than 0.5. Based on this plot, it appears that the MCED proposed in (4) may have a desirable cost-benefit for Lung and CRC cancers, although the confidence intervals for PVP for these two readouts are wide.

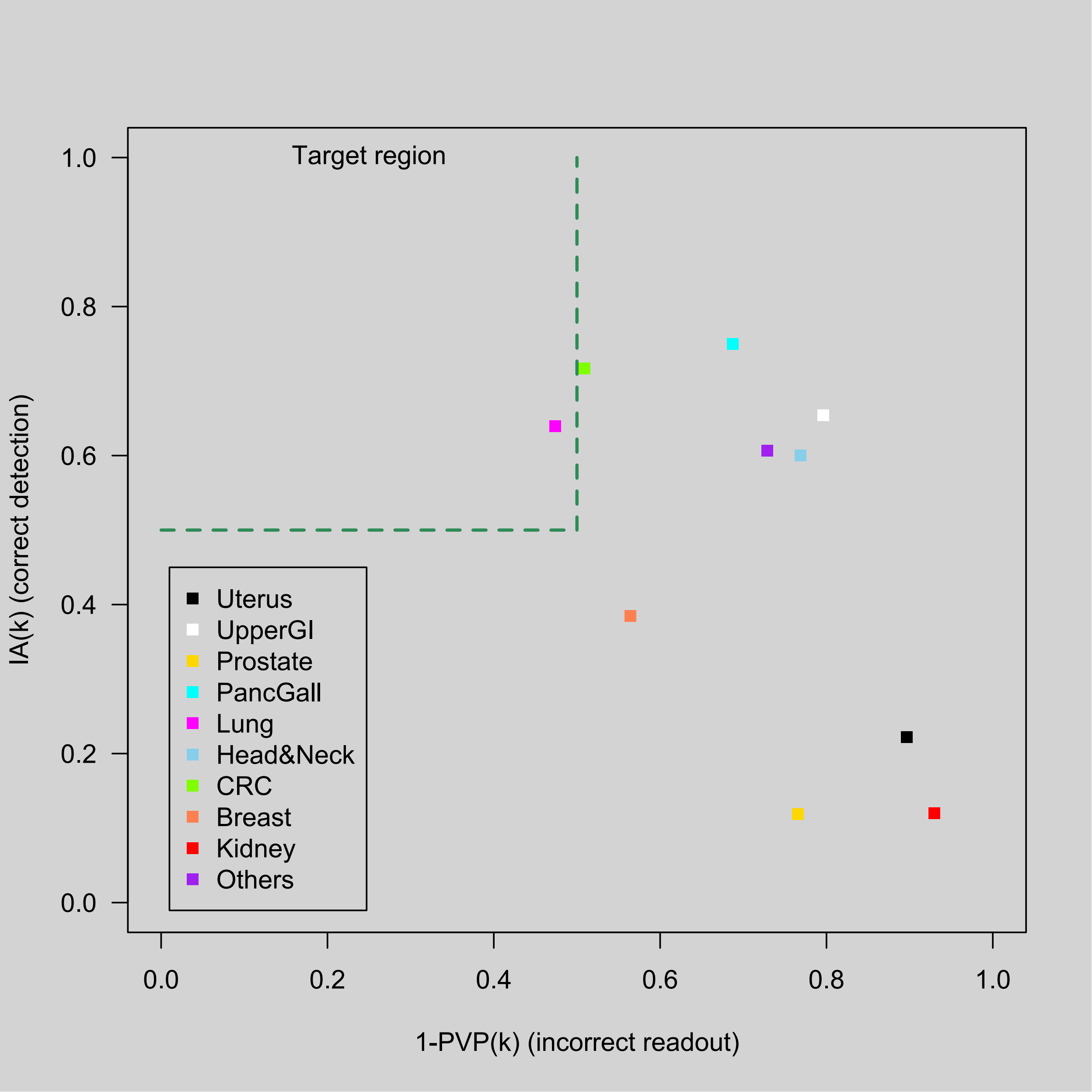

**Figure 1:** Scatter plot of empirical 1-PVP(k) (“cost”) versus corresponding IA(k) (“benefit”) for the 10 cancer types of the MCED test (4).

## 8. Discussion

The concept of testing for the presence of multiple disease types (such as multiple cancers using MCED tests) in a single blood draw is very attractive. These tests require unbiased estimation techniques and valid inference procedures for relevant performance metrics on a per-disease type level to properly inform clinical utility. Evaluating MCED tests using aggregate metrics across cancer types generally masks biological heterogeneity and cancer type-specific performance, thereby impeding a proper understanding of actualized clinical utility. In the context of a case-control study, our approach presents analytical results for Tissue-specific Predictive Performance (TSPP) that establish a framework for calculation of unbiased estimators for cancer-specific intrinsic accuracy, TOO-specific predictive value positive and the marginal test classification distribution along with corresponding valid confidence interval estimation. The methodology has the potential to assess the impact of an MCED test for informing diagnostic pathways in healthcare systems, facilitate MCED test optimization, standardize comparisons of competing tests, and inform benefit-risk analyses to inform follow-up decisions. Such capabilities can accelerate diagnosis, reduce healthcare costs, and enhance equitable access—attributes critical to regulatory and clinical adoption.

Conventional evaluations of MCED tests generally rely on three statistically disconnected metrics. One is an “overall PPV” that aggregates positive test results across all TOO readouts. This metric provides no information about the expected predictive ability of a given TOO readout; the probability of the correct case-type per given TOO readout is a crucial piece of information for clinical decision making. Another common metric is referred to as “TOO accuracy”, that calculates the overall observed proportion of correct localization among those targeted cancer case types with a positive test readout; this aggregate metric does not directly incorporate information on disease incidence and conveys no prospective actionable information because it conditions on a known cancer status and type. A third misleading metric often used in the analysis of an MCED test is a version of cancer-specific “sensitivity” that does not separate false-positive from true-positive test results. This aggregate and uninformative crude sensitivity metric does not quantify the test’s true intrinsic accuracy per cancer type. This methodologically fragmented approach often used in published results for MCED tests obscures the underlying heterogeneity in test performance and impedes a coherent assessment of test performance, thereby precluding a properly informed evaluation of the test for aiding clinical decisions. Our proposed TSPP methodology establishes unbiased estimates of cancer-specific intrinsic accuracy, TOO-specific predictive value positive and the marginal test classification distribution along with valid corresponding confidence intervals, offering a clearly focused, clinically pointed and relevant lens on meaningful and actionable information to assess the risk-benefit profiles for each disease state and test readout category.

The shortcomings become clearly evident when attempting to use information provided by conventional MCED evaluations as a basis to form a clinically actionable interpretation of test localization with diagnostic resolution on a per-patient level. Aggregate PPV and TOO accuracy do not provide a practicing clinician, with a given TOO test readout in-hand for a patient under care, the information to use the test readout to aid in clinical decision making, thereby not providing the vital information necessary for prioritizing appropriate follow-up diagnostics. An overall PPV value may in fact be constructed using crude sensitivity metrics and arise as a composite combination of these complement false-negative rates and varying cancer incidences that in general can mask an underlying skewed intrinsic accuracy profile. TOO accuracy is also non-informative for clinical patient management, not only because it is aggregate in nature but also because it conditions on those cases with a positive test readout with cancer types corresponding to the test's TOO readout categories. Since a prospective subject's cancer status is unknown at test issuance, and furthermore, if the subject is a cancer case, if the cancer type is of those targeted by the test, TOO accuracy does not convey prospective clinically actionable information for use in patient management.

The TSPP methodology introduced here presents unbiased estimation and valid inference for tissue-specific intrinsic accuracy and TOO-specific predictive value—the probability that a given positive TOO readout corresponds to the target cancer type—into a unified MCED analysis framework. These metrics assess variation in underlying cancer-type specific performance otherwise obscured by conventional aggregate metric counterparts, offering a delineated probabilistic basis for assessing the risk-benefit profile per TOO readout. Our proposed framework for targeting clinically actionable information can be used to guide decisions about whether a given result suggests a strategy of immediate invasive diagnostics or, initial evaluations with diagnostic laboratory testing or imaging for first-line workup. Unlike conventional approaches to MCED performance quantification that only provide fragmented, unfocused and distorted informational snapshots, our framework provides a cohesive, focused, and interpretable probabilistic basis for quantifying MCED performance that can directly inform clinical decision-making and an actionable follow-up strategy that aims to improve outcomes while minimizing risk.

The accuracy of a given TOO readout category can vary between early- and late-stage cancers due to biological differences in tumor burden or marker expression. Conventional metrics that report aggregate predictive value performance mask the underlying variation corresponding to stage-specific heterogeneity. For cancers with established screening programs (breast, colorectal, cervical, lung), it is critical to obtain early-stage intrinsic accuracy to inform the likelihood of timely detection and the impact on altering prognosis. A test with an apparently favorable observed aggregate PPV might perform well in assigning a positive test readout to a common late-stage cancer (e.g. colorectal), but falter in early stages or in providing the correct TOO readout, where routine screening modalities compete, rendering this aggregate metric misleading. Our methodology embeds estimation and inference for stage-stratified intrinsic accuracy metrics and furthermore breaks down TOO-specific predictive value positive into components corresponding to disease stage within a unified framework. Our proposed tools for MCED analysis can thus delineate performance gradients—e.g., an overall high colorectal TOO-PVP of say 0.85 may, for example, be decomposed into 0.35 for stages I and II and 0.50 in stages III and IV—offering a transparent basis for evaluating efficacy across the disease continuum and ensuring that screening-relevant cancers are not underserved by inflated, crude and stage-agnostic claims.

The impact on clinical practice afforded by information obtained by the appropriate analysis of MCED tests extends beyond initial screening to include the potential for shaping subsequent testing strategies, where diagnostic intervention risks must be weighed against benefits. Conventionally reported aggregate metrics do not provide information for establishing a risk-benefit guided strategy for evaluating different choices for follow-up pathways, leaving clinicians without probabilistic anchors to enable objectively informed decisions that attempt to optimize outcomes and minimize risk for the individual patient. Our MCED analysis framework provides a basis for analysis that provides required information to form strategies that aim to optimize patient outcomes by aligning intervention intensity with confidence in the reported test classification. Cancer signals resulting in a test readout category with an acceptably large corresponding predictive value positive may warrant immediate invasive diagnostics; whereas cancer signals corresponding to a test readout with a considered small predictive value positive suggest a stepwise, less invasive approach to mitigate procedural risks. For example, a colorectal cancer signal with a TOO-specific predictive value positive of 0.85 and 0.60 early-stage intrinsic accuracy could suggest direct colonoscopy—balancing a 0.15 incorrect TOO readout risk against delayed diagnosis morbidity—whereas a pancreatic signal with a predictive value positive of 0.30 might first trigger imaging, reserving biopsy for confirmed findings. This stratification, unattainable with dissociated and aggregate metrics, informs clinical actionability by quantifying diagnostic yield against procedural hazards.

The risk-benefit calculus extends to overdiagnosis—detection of indolent cancers unlikely to progress—a concern obscured by conventional metrics. Conventional aggregate metrics cannot distinguish indolent cancers from actionable diagnoses, whereas our cancer-specific TOO-PPVs highlight readouts with lower predictive value, which may reflect cancers prone to indolent behavior due to weaker molecular signals when interpreted in the proper cancer-specific setting. For example, a prostate TOO signal may frequently detect slow-growing tumors, unlike a lung signal, which typically indicates aggressive disease. This granularity enables clinicians to assess overdiagnosis risk across cancer types and stages, guiding follow-up decisions to avoid unnecessary procedures while prioritizing actionable diagnoses.

Another critical component of a test's performance is quantification of the cancer-specific false-negative and incorrect TOO classification proportions. A false-negative test result engenders false reassurance, which is particularly concerning for cancers with established screening guidelines; traditional metrics that fail to highlight this aspect may thus fail to signal when a negative result might undermine trust in routine screening programs. Incorrect TOO readout classifications can result in unnecessary and potentially harmful intervention. Conventional metrics that calculate cancer-specific crude sensitivity simply assess the likelihood of a "positive test" and ignores the actual TOO classification. Such metrics thereby mask TOO classification errors and quantify an agnostic metric of the MCED test's ability to detect the cancer type under consideration. Our MCED test analysis framework's ability to both quantify cancer-specific false-negative proportions and intrinsic accuracy measures (for the correct corresponding TOO readout) directly address the pitfalls of conventional analysis approaches by quantifying and decomposing detection failures across tumor types and stages. For example, a breast cancer false-negative proportion of 0.25 in early stages versus 0.10 overall might indicate a 25% miss rate for mammography-eligible lesions, risking patient complacency if a negative MCED test result is misinterpreted as definitive. Furthermore, a reported positive test proportion of 80% for colorectal cancer may in fact include 20% of cases that incorrectly obtained an Upper GI or Lung TOO readout; the correct intrinsic accuracy is thus 60%. The transparency and focus underlying our analysis tools enables evaluation of an MCED test's compatibility with existing screening paradigms and provides evidence to mitigate false reassurance by identifying cancer types where supplementary screening remains essential despite a negative test.

In summary, the proposed Tissue-specific Predictive Performance (TSPP) methodology is a unified analytical framework offering multiple key diagnostic metrics: cancer-specific intrinsic accuracy, Tissue-of-Origin-specific predictive value positive (PVP), and marginal test classification distributions. By bringing these metrics together into a coherent statistical model for individual tissue-specific (or cancer specific) diagnostic accuracy, it addresses critical limitations inherent in conventional evaluation methods, that obscure tissue-specific or cancer-specific performance variations, resulting in diagnostic biases and misinformation arising from use of aggregate and unfocused metrics alone.

Adopting the proposed test analysis framework can facilitate the clinical interpretability of complex, multi-category or multi-cancer diagnostic data, providing a clear and actionable probabilistic basis for decision-making. It facilitates informed risk-benefit assessments across disease (cancer) types and stages, thereby guiding follow-up clinical strategies ranging from intervention to monitoring. Furthermore, TSPP's detailed insights can improve future study design and sizing decisions for multi-disease diagnostics, while also standardizing comparisons across competing tests and diverse diagnostic platforms, whether biological assays, digital AI-based methods, or hybrid approaches. By integrating disease-specific diagnostic capability into a single, interpretable analytical structure, this methodology carries the potential to streamline diagnostic decision-making for clinicians, policymakers, scientists, and patients alike, promoting optimized, equitable, and evidence-based implementation of multi-disease tests in healthcare.

**References**

1. Ahlquist, D.A. Universal cancer screening: revolutionary, rational, and realizable. *npj Precision Onc* **2**, 1-5 (2018).
2. Putcha, G., Gutierrez, A. & Skates S. Multicancer Screening: One size does not fit all. *JCO Precision Oncology* **5**, 574-576 (2021).
3. Mercaldo, N., Lau, K. & Zhau, Z. Confidence intervals for predictive values with an emphasis to case-control studies. *Statistics in Medicine* **26**, 2170-2183 (2007).
4. Liu, M., Oxnard, G., Klein, E., Swanton, C., Seiden, M. & on behalf of the CCGA Consortium. Sensitive and specific multi-cancer detection and localization using methylation signatures in cell-free DNA. *Annals of Oncology* **31**, 745-759 (2020).
5. Agresti, A. & Gottard, A. Comment: Randomized confidence intervals and the Mid-*P* approach. *Statistical Science* **20**, 367-371 (2005).
6. Goodman, L.A. Interactions in multi-dimensional contingency tables. *Annals of Mathematical Statistics* **35**, 632-646 (1964).
7. Goodman, L.A. The multivariate analysis of qualitative data: Interaction among multiple classifications. *Journal of the American Statistical Association* **65**, 226-256 (1970).
8. Goodman, L.A. The analysis of multidimensional contingency tables: Stepwise procedures and direct estimation methods for building models for multiple classifications. *Technometrics* **13**, 33-61 (1971).
9. SEER*Explorer: An interactive website for SEER cancer statistics [Internet]. Surveillance Research Program, National Cancer Institute; 2024 Apr 17. Available from: https://seer.cancer.gov/statistics-network/explorer/. Data source(s): SEER Incidence Data, November 2023 Submission (1975-2021), SEER 22 registries.

## Competing interests

The authors declare no competing interests.

## Author contributions

All authors conceived of and designed the study. A.G.D wrote the original draft of the manuscript, curated all data and wrote all computer code for analysis and visualization. All authors reviewed and edited the final draft of the manuscript.

## Ethics

Because only publicly available and simulated data were used in this manuscript, there was no ethical approval of any kind that was required.

## Appendix

$P(n_{j+} > 0)$ is considered fixed when constructing variance estimators.

### A1: Components of $\boldsymbol{U}'(\boldsymbol{\varphi_k})$, $k = 1, \ldots, K$.

(1) $\frac{\partial U(\boldsymbol{\varphi_k})}{\partial P(T_k|D_0)} = -\{1 - P(D)\}/R_U(\varphi_k)$

(2) $\frac{\partial U(\boldsymbol{\varphi_k})}{\partial P(T_k|D_k)} = 1/P(T_k \mid D_k)$

(3) For $j = 1, \ldots, J, j \neq k$:

$$\frac{\partial U(\boldsymbol{\varphi_k})}{\partial P(T_k|D_j)} = -\{I(j \neq k)P\left(D_j \mid D\right)P(D)\}/R_U(\varphi_k)$$

(4) For $k < J$:

$$\frac{\partial U(\boldsymbol{\varphi_k})}{\partial P(D_k|D)} = \frac{1}{P(D_k|D)} + \frac{I(k \neq J)P(D)P(T_k|D_J)}{R_U(\varphi_k)}$$

(5) For $k < J, j = 1, \ldots, J-1, j \neq k$:

$$\frac{\partial U(\boldsymbol{\varphi_k})}{\partial P(D_j|D)} = -P(D)\left\{I(k \neq j)P\left(T_k \mid D_j\right) - I(k \neq J)P\left(T_k \mid D_J\right)\right\}/R_U(\boldsymbol{\varphi_k})$$

(6) For $k = J, j = 1, \ldots, J-1$:

$$\frac{\partial U(\boldsymbol{\varphi_k})}{\partial P(D_j|D)} = \frac{-1}{\left\{1 - \sum_{\ell=1}^{J-1} P\left(D_\ell|D\right)\right\}} - \frac{I(k \neq j)P(D)P(T_k|D_j)}{R_U(\varphi_k)}$$

**A2: Components of $\boldsymbol{W'(\varphi_k)}, k = 0, \dots, K$.**

Define $R_W(\boldsymbol{\varphi_k}) = P(D)\sum_{j=1}^{J-1} P(T_k \mid D_j)P(D_j \mid D) + P(D)P(T_k \mid D_J)\{1 - \sum_{j=1}^{J-1} P\big(D_j \mid D\big)\}$

(1) $\frac{\partial W(\boldsymbol{\varphi_k})}{\partial P(T_k \mid D_0)} = 1/P(T_k \mid D_0)$

(2) For $j = 1, \dots, J-1$:

$\frac{\partial W(\boldsymbol{\varphi_k})}{\partial P(T_k \mid D_j)} = -P(D)P(D_j \mid D)/R_W(\boldsymbol{\varphi_k})$

(3) For $j = J$:

$\frac{\partial W(\boldsymbol{\varphi_k})}{\partial P(T_k \mid D_J)} = -P(D)\{1 - \sum_{\ell=1}^{J-1} P(D_\ell \mid D)\}/R_W(\boldsymbol{\varphi_k})$

(4) For $j = 1, \dots, J-1$:

$\frac{\partial W(\boldsymbol{\varphi_k})}{\partial P(D_j \mid D)} = -P(D)\{P\big(T_k \mid D_j\big) - P(T_k \mid D_J)\}/R_W(\boldsymbol{\varphi_k})$

**A3: Components of $(2J \times 2J)$ matrix $\boldsymbol{V(\widehat{\varphi}_k - \varphi_k)}, k = 0, \dots, K$.**

$$\boldsymbol{V}(\boldsymbol{\widehat{\varphi}_k} - \boldsymbol{\varphi_k}) = \begin{pmatrix} N_0^{-1}P(T_k \mid D_0)\{1 - P(T_k \mid D_0)\} & 0 & \cdots & 0 \\ 0 & & \boldsymbol{V}_1 & \boldsymbol{V}_2 \\ \vdots & & & \\ 0 & & \boldsymbol{V}_2^t & \boldsymbol{V}_3 \end{pmatrix}$$

$\boldsymbol{V}_1 = \mathrm{diag}(\sigma_{1k}^2, \cdots, \sigma_{Jk}^2)$, a$(J \times J)$ matrix, since $Cov\left(\frac{\tilde{A}_{jk}}{P(n_{j+}>0)}, \frac{\tilde{A}_{\ell k}}{P(n_{\ell +}>0)}\right) = 0, (j \neq \ell)$, under the assumptions:

(1) $E\left(Y_i^{(jk)} \mid Y_i^{(\ell k)}, n_{j+}, n_{\ell +}\right) = E(Y_i^{(jk)})$ and $E\left(Y_i^{(jk)} \mid n_{j+}, n_{\ell +}\right) = E(Y_i^{(jk)})$ these follow from the definition of $\tilde{A}_{jk}$ as a compound random variable.

(2) $P\{(n_{j+} > 0) \mid (n_{\ell +} > 0)\} = P\{(n_{j+} > 0)\}$, this is a reasonable assumption when similar disease types are grouped when defining $D_1 \cdots D_J$.

$$\boldsymbol{V}_2 = \begin{pmatrix} \eta_{11}^{(k)} & 0 & \cdots & 0 \\ 0 & & & 0 \\ \vdots & & \ddots & \vdots \\ 0 & & \cdots & \eta_{(J-1)(J-1)}^{(k)} \\ 0 & & \cdots & 0 \end{pmatrix}$$, a $(J \times (J-1))$ matrix

$\eta_{j\ell}^{(k)} = Cov\left(\frac{\tilde{A}_{jk}}{P(n_{j+} > 0)}, \frac{n_{\ell+}}{N_1}\right), j = 1, \ldots, J, \ell = 1, \ldots, J - 1.$

Under the assumption $E\left(Y_i^{(jk)} \mid n_{j+}, n_{\ell+}\right) = E(Y_i^{(jk)})$,

when $j = \ell$:

$$\eta_{j\ell}^{(k)} = P(T_k \mid D_j)\left\{\frac{1}{N_1}E(n_{j+} \mid n_{j+} > 0) - P(D_j \mid D)\right\}$$

when $j \neq \ell$:

$$\eta_{j\ell}^{(k)} = P(T_k \mid D_j)\left\{\frac{1}{N_1}E(n_{\ell+} \mid n_{j+} > 0) - P(D_\ell \mid D)\right\}$$

= 0 when $E(n_{\ell+} \mid n_{j+} > 0) = E(n_{\ell+})$.

$\boldsymbol{V}_3$ is the $(J-1) \times (J-1)$ multinomial variance covariance matrix. Define the column vector $\boldsymbol{v} = \{P(D_1 \mid D), \ldots, P(D_{J-1} \mid D)\}^t$

$$\boldsymbol{V}_3 = N_1^{-1}\{\text{diag}(\boldsymbol{v}) - \boldsymbol{v}\boldsymbol{v}^{\boldsymbol{t}}\}.$$